\documentclass[twocolumn,showpacs,prl,amsmath,amssymb,aps]{revtex4}
\usepackage{graphicx}
\usepackage{amsthm}
\usepackage{epsfig}
\usepackage{mathrsfs}
\def\<{\langle}\def\>{\rangle}
\def\Tr{\operatorname{Tr}}

\def\Reals{\mathbb R}
\def\spc#1{\mathcal{#1}}

\def\Bndd#1,#2{\mathcal{B}(#1,#2)}

\begin{document}
\title{Optimal estimation of squeezing}
\author{G.~Chiribella} 
\author{G.~M.~D'Ariano} 
\altaffiliation[Also
at ]{Center for Photonic Communication and Computing, Department of
Electrical and Computer Engineering, Northwestern University,
Evanston, IL 60208} 
\affiliation{QUIT Quantum Information Theory Group}
\homepage{http://www.qubit.it}
\affiliation{Dipartimento di Fisica
``A. Volta'', University of Pavia, via A. 
Bassi 6, I-27100 Pavia, Italy} 
\author{M.~F.~Sacchi} \altaffiliation[Also
at ]{Consorzio Nazionale Interuniversitario per la 
Struttura della Materia}
\affiliation{QUIT Quantum Information Theory Group}
\homepage{http://www.qubit.it}
\affiliation{CNR - Istituto Nazionale per la Fisica della Materia, 
via A. Bassi 6, I-27100 Pavia, Italy} 

\date{\today}

\begin{abstract}
  We present the optimal estimation of an unknown squeezing
  transformation of the radiation field. The optimal estimation is
  unbiased and is obtained by properly considering the degeneracy of
  the squeezing operator.  For coherent input states, the r.m.s. of
  the estimation scales as $(2\sqrt {\bar n})^{-1}$ versus the average
  photon number $\bar n$, while it can be enhanced to $(2\bar n)^{-1}$
  by using displaced squeezed states.
\end{abstract}

\date{\today}

\maketitle Squeezed states are characterized by a phase-dependent
redistribution of quantum fluctuations such that the dispersion in one
of the two quadrature components of the field is reduced below the
level set by the symmetric distribution of the vacuum state or a
coherent state \cite{spis}. Such a property has been used to raise
the sensitivity beyond the standard quantum limit \cite{bach,treps}
and to enhance interferometric \cite{interf} and absorption
measurements \cite{abs}, along with optical imaging applications 
\cite{kol,treps}.
\par Even though squeezed states have been studied extensively during the
last three decades, the attention to the problem of estimating an
unknown squeezing transformation is relatively recent, and very few
results are known about it.  The first attempt to quantify the
accuracy limits imposed by quantum mechanics was presented in Ref.
\cite{milb}, in the case of squeezing in a fixed direction. Here, the
squeezing transformations form a one-parameter group, and the
estimation problem is closely similar to the problem of phase
estimation \cite{HolevoPhase,pomph} (for this reason the name
\emph{hyperbolic phase estimation} has been also used).  The basic
idea underlying the estimation strategy is to find a measurement that
projects the quantum state on the vectors that are canonically
conjugated via Fourier transform to the eigenstates of the squeezing
generator.  However, as we will show in this Letter, the scheme of
Ref.  \cite{milb} is not optimal and is biased: neither the mean value
nor the most likely one in the probability distribution coincide with
the true value of the squeezing parameter.  In other words, the
estimation is biased, and the presence of such a bias suggests that
the proposed scheme is not optimal.

More recently, the estimation of squeezing has been considered in
connection with cloning \cite{HayashiCloning}. In this case the
unknown squeezing is estimated from a number of identical copies of
the same unknown squeezed state. However, this approach does not work
when only a single copy is available, and the problem of the bias and
the optimality of the estimation is left open.

In this Letter, we will present the optimal estimation of an unknown
squeezing transformation in a given direction, acting on an arbitrary
state of the radiation field. This problem is the optimal estimation
of squeezing in an experimental situation where a degenerate
parametric amplifier is pumped by a strong coherent field with a fixed
phase relation with the state to be amplified.

We will show that the optimal measurement is unbiased, provided that
one properly takes into account the degeneracy of the squeezing
operator.  Due to such a degeneracy, the Fourier transform of the
eigenstates of the squeezing operator is not uniquely defined, and, in
order to obtain the best estimation strategy, one has to perform an
optimization similar to that of phase estimation with degeneracy
\cite{pomph}.  Accordingly, the optimal estimation of squeezing
depends on the chosen initial state of the radiation field.  Also, the
optimization performed here is analogous to that of Ref.
\cite{Refframe} in the case of estimation of rotations, namely it
properly takes into account the equivalent representations of the
group of parameters.  In fact, the degeneracy of the squeezing
operator corresponds to the presence of equivalent representations of
the related one-parameter group.

We will derive our results in the framework of quantum estimation
theory \cite{Helstrom,Holevo}, upon defining optimality as the
minimization of the expected value of a given cost function, which
quantifies the deviation of the estimated parameter from the true one.
According to the minimax approach, the optimal estimation strategy
will be the one that minimizes the maximum of the expected cost over
all possible true values of the unknown squeezing parameter. In
analogy with the class of cost functions introduced by Holevo
\cite{HolevoPhase,Holevo} for the problem of phase estimation, we
introduce here a class of cost functions including a large number of
optimality criteria, such as maximum likelihood, and maximum fidelity.
We will show that our estimation strategy is optimal according to any
function in such a class.

In the following, we consider a single-mode radiation field with
bosonic operators $a$ and $a^\dag $, satisfying the canonical
commutation relations $[a, a^\dag ]=1$.  The squeezing operator is
defined as follows
\begin{eqnarray}
S(r)=\exp \left[ \frac r2 \left ( a^{\dag 2} -a^2 \right) \right ]\;,
\end{eqnarray}
where $r$ is a real parameter. Given a pure state $|\psi \rangle $ of
the radiation field, we want to find the optimal measurement that
allows one to estimate the parameter $r$ in the transformation $ |\psi
\rangle \longrightarrow S(r)|\psi \rangle$.  In the quadrature
representation $\psi (x) = \langle x |\psi \rangle $, where $|x
\rangle $ denotes the Dirac-normalized eigenvector of the quadrature
operator $X=(a +a^\dag )/2$, the effect of squeezing on the
wavefunction is given by $\psi (x) \longrightarrow e^{-r/2} \psi
(e^{-r}x)$.

The squeezing operator can be written as $S(r)=e^{- i r K}$, where $K$
is the Hermitian operator $K = i (a^{\dag 2} - a^2)/2$, that generates
the one-parameter group of squeezing transformations.  The spectrum of
the generator $K$ is the whole real line, and the eigenvalue equation
reads
\begin{eqnarray} 
 K |\mu ,s \rangle 
=\mu |\mu ,s\rangle \;,\label{eig}
\end{eqnarray}  
where $\mu \in \Reals$ is the eigenvalue, and $s$ is a degeneracy
index with two possible values $\pm 1$. The explicit expression of the
generalized eigenvectors of $K$ in the quadrature representation is given by 
\cite{boll}
\begin{eqnarray} \label{BOx}
\langle x|\mu , s \rangle
=\frac{1}{\sqrt{2\pi}}\,|x|^{i\mu-\frac 12}~\theta(s x)\;,
\end{eqnarray}  
where $\theta(x)$ is the Heaviside step-function [$\theta(x)=1$ for
$x>0$, $\theta(x)=0$ for $x<0$].  The vectors $|\mu, s\>$ are
orthogonal in the Dirac sense, namely $\langle \mu,r |\nu, s \rangle
=\delta _{rs}~\delta(\mu-\nu)$, and provide the resolution of the
identity
\begin{equation}\label{Completeness}
\int_{-\infty}^{+\infty} d \mu~ \Pi_{\mu} = \openone~,
\end{equation}
where $\Pi_{\mu}= \sum_{s= \pm 1} ~|\mu, s\>\<\mu, s|$ is the
projector onto the eigenspace of $K$ corresponding to the eigenvalue
$\mu$.

Let us denote by $\spc H_{\mu}$ the two-dimensional vector space
spanned by $|\mu, \pm 1\>$. In this complex vector space, we can
consider the usual scalar product and the corresponding norm, namely
if $|v_{\mu}\>=\sum_{s = \pm 1}v_s^{\mu} ~|\mu,s\>$ is an element of
$\spc H_{\mu}$, then its norm is $|| |v_{\mu}\>||= \left(\sum_{s= \pm
    1} ~|v_s^{\mu}|^2\right)^{1/2}$.  Using the completeness relation
(\ref{Completeness}), we can write any pure state $|\psi\> \in \spc H$
as
\begin{equation}\label{StateDecomp}
|\psi\> = \int_{-\infty}^{+\infty} d \mu~ c_{\mu} |\psi_{\mu}\>~,
\end{equation}  
where $c_{\mu} =|| \Pi_{\mu}|\psi\>||$,  and 
\begin{equation}\label{Projection}
|\psi_{\mu}\>= \frac{\Pi_{\mu} |\psi\>}{||\Pi_{\mu} |\psi\>||}
\end{equation}
is the normalized projection of $|\psi\>$ onto $\spc H_{\mu}$. 
The representation of a state as in Eq. (\ref{StateDecomp})
corresponds to the fact that the Hilbert space $\spc H$ can be
decomposed as a \emph{direct integral} $\spc H =
\int_{-\infty}^{+\infty} d \mu~ \spc H_{\mu}$.  In this
representation the effect of a squeezing transformation is given by
\begin{equation}\label{SqueezPsi}
S(r) |\psi\>= \int_{-\infty}^{+\infty} 
d \mu ~ c_{\mu} e^{ - i r \mu }~|\psi_{\mu}\>~,
\end{equation}
i.e. the squeezing operator introduces a different phase shift in any
space $\spc H_{\mu}$.  Notice that the states (\ref{SqueezPsi}) all
lie in the subspace
\begin{equation}\label{HPsi}
\spc H_{\psi}= 
\left \{|v\>= \int_{-\infty}^{+\infty} 
d \mu~ v_{\mu} |\psi_{\mu}\>~\left|~  v_{\mu} \in L^2(\Reals) \right. \right \}~.
\end{equation}
The problem of squeezing estimation in the representation
(\ref{SqueezPsi}) becomes formally equivalent to the problem of phase
estimation.

In order to optimize the estimation of squeezing, we describe the
estimation procedure with a \emph{positive operator valued measure}
(POVM) $P(\hat r)$. The probability distribution of estimating $\hat
r$ when the true value of squeezing is $r$ is then given by $p(\hat r|
r) = \Tr[ P(\hat r) S_r \rho S_r^{\dag}] $.  The optimality criterion
is specified in terms of a \emph{cost function} $c(\hat r-r)$, that
quantifies the cost of estimating $\hat r$ when the true value is $r$.
Once a cost function has been fixed, the optimal measurement is
defined in the minimax approach as the one that minimizes the quantity
\begin{equation}\label{MaxCost}
\bar c = \max_{r \in \Reals} \left \{ \int_{-\infty}^{+\infty}
d \hat r~  p(\hat r|r) c(\hat r-r)\right\}~,
\end{equation} 
namely the maximum of the expected cost over all possible true values.
Generalizing the class of cost functions introduced by Holevo for
phase estimation \cite{Holevo}, we consider here cost functions of the
form
\begin{equation}\label{HolevoClass}
c(r)= \int_{0}^{+\infty} d \mu ~ a_{\mu} \cos ( \mu r)~,
\end{equation}
where $a_{\mu} \le 0$ for any $\mu >0$. This class contains a
large number of optimality criteria, such as the maximum likelihood
$c_{ML}(r)= -\delta (r)$, and the maximum fidelity $c_F(r)=1-
|\<\psi|S(r)|\psi\>|^2$.

Due to the group symmetry of the problem, instead of searching among
all possible measurements for optimization, one can restrict attention
to the class of \emph{covariant} measurements \cite{Holevo}, which are
described by POVMs of the form $P(\hat r)= S(\hat r) \xi S(\hat
r)^{\dag}$, with $\xi \ge 0$ such that
\begin{equation}
\int_{-\infty}^{+\infty}~ d r~ S(r) \xi S(r)^{\dag} = \openone~.
\end{equation} 
The probability distribution $p(\hat r|r)$ related to a covariant
measurement will depend only on the difference $\hat r-r$, and this
means that the estimation is equally good for any possible value of
the unknown squeezing \cite{nota}. 

The optimization of the covariant measurement for any cost function in
the class (\ref{HolevoClass}) can be obtained as in the case of phase
estimation with degeneracy \cite{pomph}. The optimal covariant POVM is
then given by
\begin{equation}\label{OptPovm}
P(r) = |\eta (r)\rangle \langle \eta (r)|\;,
\end{equation}
where 
\begin{eqnarray}\label{EtaOpt}
|\eta (r) \rangle = \int _{-\infty }^{+\infty } \frac{d\mu}{\sqrt{2\pi}} \,
e^{- i r \mu }\,|\psi_{\mu} \rangle \;.
\end{eqnarray}   
Notice the correspondence of $|\eta (r) \rangle $ with the vectors
$|e(\varphi )\rangle = \sum _{n=0}^\infty \frac{e^{in
    \varphi}}{\sqrt{2\pi}} |n \rangle $ that arise in the context of
optimal phase estimation (here $|n\>$ are the non-degenerate
eigenvectors of the photon number operator $a^{\dag}a$). The vectors
$|\eta(r)\>$ are orthogonal in the Dirac sense, namely the optimal
POVM is a von Neumann measurement. The projection $|\psi _\mu \rangle
$ of Eq.  (\ref{Projection}) in the expression of $|\eta (r) \rangle $
makes the optimal measurement depend on the input state $|\psi \rangle
$. Accordingly, one obtains non-commuting observables, corresponding to
different input states.  The normalization of the POVM (\ref{OptPovm})
can be easily checked, since $\int _{-\infty }^{+\infty } dr~ P(r) =
\openone_{{\psi}}$~, where $\openone_{\psi}$ is  the identity in the
subspace ${\cal H}_\psi$ defined in Eq.  (\ref{HPsi}). Clearly, the
$P(r)$ can be arbitrarily completed to the whole Hilbert space,
without affecting the probability distribution of the outcomes.

Using Eq. (\ref{Projection}), the optimal probability distribution for
an input state $|\psi \rangle $ is given by
\begin{eqnarray}
p(\hat r|r) &=&\langle \psi |S(r)^{\dag}~ P(\hat r)~ S(r)|\psi \rangle
\nonumber \\ &=&\frac {1}{2\pi} \left | \int
_{-\infty }^{+\infty } d\mu \, e^{- i (\hat r-r) \mu }\,\sqrt {\langle \psi |\Pi
_\mu |\psi \rangle }\,\right |^2 \;.\label{pdir}
\end{eqnarray}
Since the probability distribution depends only on the difference
$\hat r -r$, from now on we will write $p(\hat r-r)$ instead of
$p(\hat r|r)$.  Representing the projection $\Pi _\mu $ as $\Pi _\mu =
\int _{-\infty }^{+\infty } \frac {d \lambda }{2 \pi }\, e^{i \lambda
  (\mu -K )} $, the probability distribution of Eq. (\ref{pdir}) can
be rewritten as
\begin{eqnarray}\label{pdir'}
p(r)=\left | \int _{-\infty }^{+\infty } \frac {d \mu }{2\pi }\, e^{-i r
\mu }\, \sqrt{\int _{-\infty }^{+\infty } d \lambda \, e^{i \lambda
\mu } \,\langle \psi |S(\lambda )|\psi \rangle }\, \right
|^2\,.
\end{eqnarray}

The optimal measurement (\ref{OptPovm}) can be compared with that given in
Ref. \cite{milb}, which is described in our notation by the POVM 
\begin{eqnarray}\label{MilbPovm}
\tilde P(r) = \sum_{s= \pm 1} |\eta_s (r)\>\<\eta_s (r)|~, 
\end{eqnarray}
where 
\begin{equation}\label{EtaMilb}
|\eta_{s}(r)\>= \int _{-\infty}^{+\infty} 
\frac{d \mu}{\sqrt{2\pi}} ~ e^{- ir\mu} |\mu, s \>~.
\end{equation} 
Using Eq. (\ref{BOx}), it is easy to see that $|\eta_{\pm}(r)\>$ are
eigenvectors of the quadrature $X$ corresponding to the eigenvalues
$\pm e^r$, and hence the POVM (\ref{MilbPovm}) corresponds to
measuring the observable $\ln |X|$, {\em independently} of the input
state. The measurement $\tilde P(r)$ is not optimal, and gives a
biased probability distribution, namely the average value of the
estimated parameter does not coincide with the true value, and also
the most likely value in the probability distribution is not the true
one (see, e.g., the asymmetric probability distribution for the vacuum
state in Fig. 1). Such drawbacks do not occur in the optimal
probability distribution (\ref{pdir}). Notice also that the
measurement $\tilde P(r)$ is ``rank-two'' in the subspace $\spc
H_{\psi}$ of interest, while the optimal measurement is ``rank-one''.
The differences between the two measurements can be understood
intuitively as follows. Essentially, both POVMs are based on the
Fourier transform of the eigenvectors of the operator $K$. However,
since the Fourier transform is not uniquely defined due to the
degeneracy of $K$, one should optimize it versus the input state.

\begin{figure}[htb]
\begin{center}
\includegraphics[scale=1]{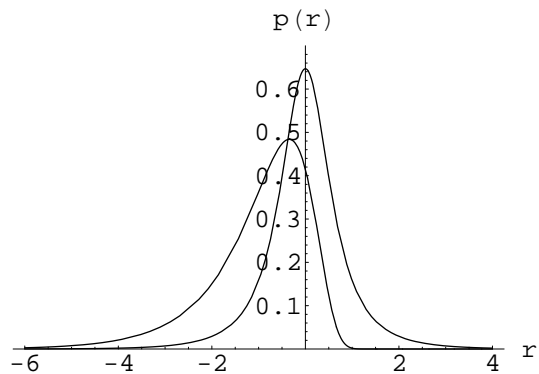}
\caption{
  Probability distributions for the estimation of squeezing on a
  vacuum input state. The asymmetric distribution comes from the
  suboptimal measurement of Ref. \cite{milb} in Eq.  (\ref{MilbPovm}).
  The symmetric distribution corresponds to the optimal measurement of
  Eq.  (\ref{OptPovm}).}
\label{f:fig1}
\end{center}
\end{figure}

\begin{figure}[htb]
\begin{center}
\includegraphics[scale=1]{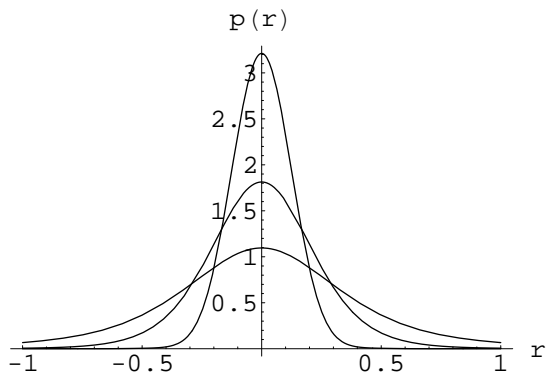}
\caption{Optimal probability distribution of squeezing 
  for input coherent states. The distribution becomes sharper for
  increasing values of the coherent-state amplitude ($\alpha =1,\,2,\,4$.)}
\label{f:fig2}
\end{center}
\end{figure}

In the case of a coherent input state $| \alpha \rangle $, the
probability distribution (\ref{pdir'}) can be specified as follows
\begin{eqnarray}\label{pcoh}
&&p(r)=
e^{-|\alpha |^2}  \left |\int _{-\infty
  }^{+\infty } \frac {d \mu }{2\pi } 
\,e^{- i \mu r }  \right. \times \\&  & 
\left. \sqrt{\int
    _{-\infty }^{+\infty } \frac{d \lambda }{\sqrt {\cosh \lambda }}\,
    e^{i \lambda \mu } \,e^{\frac 12 \tanh \lambda (\alpha ^{*2}-\alpha ^{2})}
\,e^{\frac{|\alpha |^2
    }{\cosh \lambda }}} \; \right |^2  \;\nonumber \label{pra} 
\end{eqnarray}
This probability distribution has been plotted for increasing real
values of $\alpha$ in Fig. \ref{f:fig2}, where one 
can easily observe the corresponding improvement in the estimation.

For large values of $|\alpha |$, from Eq. (\ref{pra}) one obtains
asymptotically the Gaussian distribution
\begin{eqnarray}
p(r) = \sqrt{\frac{2 |\alpha |^2}{\pi }}\,e^{-2 |\alpha |^2 r^2}\;,
\end{eqnarray}
that provides a r.m.s error on the estimation of $r$ as $\Delta r
=1/(2 \sqrt{\bar n})$, where $\bar n=|\alpha|^2$ is the mean photon
number. This scaling can be improved to $\Delta r = 1/(2 \bar n)$ by
using displaced squeezed states $|\alpha,z\> = D(\alpha) S(z) |0\>$,
with $\alpha,z \in \Reals$. In fact, from the relation $D (\alpha )
S(z) = S(z) D(\alpha e^{-z})$, the probability distribution $p(r)$ is
given by Eq. (\ref{pcoh}) just by replacing $\alpha $ with $\alpha
e^{-z}$. In the asymptotic limit of large number of photons $\bar n=
|\alpha|^2 + \sinh ^2 z$, the minimization of the r.m.s.
gives the optimal scaling $\Delta r = 1/(2 \bar n)$, for $\alpha=
\sqrt{\bar n/2}$ and $z=- 1/2 \ln (2\bar n)$, and this corresponds to
approximate the eigenvectors of the quadrature operator $X$.

In the asymptotic regime, the optimal performance can be achieved
simply by measuring the quadrature $X$ and estimating $\hat r = \ln
|x/\alpha|$, in correspondence to the outcome $x$.
However, it is important to stress that homodyne measurement becomes
optimal only for particular input states and in the asymptotic limit
of large energy, while for finite energy the optimal measurement is
described by the POVM in Eq. (\ref{OptPovm}).

In conclusion, we presented the covariant measurement for estimating
the squeezing that is optimal for a large class of figure of merit.
The optimal detection is given by a suitable Fourier transform of the
eigenstates of the generator of squeezing. In fact, due to the
degeneracy of the squeezing operator, there is a freedom in choosing
how to perform the Fourier transform, and the choice must be optimized
according to the input state. Hence, for different input states one
has different optimal estimations corresponding to different
observables. The optimal measurement leads to an unbiased estimation,
and the outcome of the measurement that is most likely to be obtained
coincides with the true value of the unknown squeezing.  For coherent
input states the r.m.s. error scales as $1/(2\sqrt{\bar n})$ with the
number of photons, while for displaced squeezed states one achieves
$1/(2 \bar n)$ scaling.  In the asymptotic regime, such a scaling can
be obtained experimentally by homodyne measurement.  The presented
scheme applies to the problem of optimal characterization of nondegenerate
parametric amplifiers.

\acknowledgments We acknowledge M. Hayashi for suggesting Ref.
\cite{OzawaUnpublished}. This work has been supported by Ministero
Italiano dell'Universit\`a e della Ricerca (MIUR) through FIRB (bando
2001) and PRIN 2005.


\begin{thebibliography}{99}
\bibitem{nota}The proof the in the minimax problem the optimal
  measurement can be searched among the covariant ones holds both for
  compact \cite{Holevo} and non-compact \cite{OzawaUnpublished} group.
\end{thebibliography}
\end{document}